 \documentclass[final,5p,times,twocolumn]{elsarticle}

\usepackage{graphics}
\usepackage{multirow}
\usepackage{amssymb}
\usepackage{amsmath}
\usepackage{mathrsfs}
\usepackage{xspace}

\journal{Nuclear Instruments and Methods A}
\begin{document}
\begin{frontmatter}

\title{Development Toward a Ground-Based Interferometric Phased Array for 
Radio Detection of High Energy Neutrinos}
\author[a,b]{J.~Avva}
\author[c,b]{K.~Bechtol}
\author[d]{T.~Chesebro}
\author[e]{L.~Cremonesi}
\author[b]{C.~Deaconu}
\author[e]{A.~Gupta}
\author[f,b]{A.~Ludwig}
\author[g]{W.~Messino}
\author[h]{C.~Miki}
\author[e]{R.~Nichol}
\author[b]{E.~Oberla}
\author[b]{M.~Ransom}
\author[i]{A.~Romero-Wolf}
\author[d]{D.~Saltzberg}
\author[d]{C.~Schlupf}
\author[j,b]{N.~Shipp}
\author[h]{G.~Varner}
\author[f,k,b]{A.~G.~Vieregg}
\author[l,d]{S.~A.~Wissel}

\address[a]{Dept. of Physics, University of California Berkeley, Berkeley, CA 94720, USA}
\address[b]{Kavli Institute for Cosmological Physics, University of Chicago, Chicago, IL 60637, USA}
\address[c]{Wisconsin IceCube Particle Astrophysics Center, University of Wisconsin-Madison, Madison, WI 53703, USA}
\address[d]{Dept. of Physics and Astronomy, University of California Los Angeles, Los Angeles, CA 90095, USA}
\address[e]{Dept. of Physics and Astronomy, University College London, London, United Kingdom}
\address[f]{Dept. of Physics, University of Chicago, Chicago, IL 60637, USA}
\address[g]{Electrical Engineering Dept., California Polytechnic State University, San Luis Obispo, CA 93407, USA}
\address[h]{Dept. of Physics and Astronomy, University of Hawaii, Manoa, HI 96822, USA}
\address[i]{Jet Propulsion Laboratory, California Institute of Technology, Pasadena, CA 91109, USA}
\address[j]{Dept. of Astronomy and Astrophysics, University of Chicago, Chicago, IL 60637, USA}
\address[k]{Enrico Fermi Institute, University of Chicago, Chicago, IL 60637, USA}
\address[l]{Physics Dept., California Polytechnic State University, San Luis Obispo, CA 93407, USA}


\begin{abstract}
The in-ice radio interferometric phased array technique for detection of high energy neutrinos 
looks for Askaryan emission from neutrinos interacting in large volumes of glacial ice, and is being 
developed as a way to achieve a low energy threshold and a large effective volume at high energies.
The technique is based on coherently summing the impulsive Askaryan signal from multiple antennas, which
increases the signal-to-noise ratio for weak signals.
We report here on measurements
and a simulation of thermal noise correlations between
nearby antennas, beamforming of impulsive signals, 
and a measurement of the expected improvement in trigger efficiency through the phased array technique.
We also discuss the noise environment observed with an analog phased array at Summit Station, Greenland,
a possible site for an interferometric phased array for radio detection of high energy neutrinos.

\end{abstract}


\end{frontmatter}
\section{Introduction} 
\label{sec:intro}
In recent years, the IceCube experiment has detected a 
population of astrophysical neutrinos with energies up to $\sim10$~PeV~\cite{bigBird,bertErnie}.  
The sources of these astrophysical 
neutrinos remain a mystery, their spectral index remains uncertain, and although
there is no evidence for a spectral cutoff, 
the behavior at higher energies remains unknown~\cite{icecubeGlobal}.
In addition to the astrophysical population discovered by IceCube, there is a separate 
population of cosmogenic ultra-high energy (UHE) neutrinos ($E>10^{17}$~eV), created as a byproduct 
of the GZK process (the interaction of UHE cosmic rays with the cosmic microwave background), 
that awaits discovery~\cite{g,zk,beresinsky_1969_cosmogenic}.
The twin science goals of following up on the IceCube measurement of astrophysical neutrinos at and 
above PeV~energies,
and discovering the highest energy cosmogenic neutrinos drive the design of developing and proposed 
experiments that aim to detect high energy neutrinos.  

One promising method for detection of high energy
neutrinos is via the Askaryan effect: 
the coherent, impulsive radio emission from electromagnetic showers induced by neutrinos in a 
dielectric~\cite{askaryan}.  At long wavelengths (frequency less than a few~GHz), the emission is coherent,
so for high energy showers, the long-wavelength radio emission dominates.  
A large volume 
of a dielectric material with a long radio attenuation length ($L_\alpha \sim 1$~km), such as glacial
ice, is required to detect a significant rate of high energy astrophysical and cosmogenic neutrinos.

There are a variety of current and proposed experiments that search for Askaryan emission from 
high energy neutrino showers.  The ANITA high altitude balloon experiment currently holds the 
best constraints on the flux of neutrinos 
above $10^{19.5}$~eV, and the proposed balloon-borne EVA experiment is a novel way 
to improve sensitivity at these highest energies~\cite{anita2,eva}. 
The ARA and ARIANNA experiments, ground-based 
radio arrays in early stages of development each with a small number of stations deployed in Antarctica,
have energy thresholds $\gtrapprox 100$~PeV, probing the heart of the cosmogenic neutrino 
regime~\cite{ara2015,arianna_2015}.

The concept for an in-ice radio interferometric phased array for detection
of high energy neutrinos was introduced in Reference~\cite{phasedArray} and is being explored as a way to push 
the energy threshold of radio detection down to the PeV scale while increasing the achievable 
effective volume at the highest energies.  
Interferometric techniques have been extensively used in radio 
astronomy~(for a review, see~\cite{thompson}) to image radio sources, and here we apply an interferometric
technique to improve sensitivity to broadband, impulsive radio signals.  Rather than imaging,
we are interested in achieving high instantaneous sensitivity to a large solid angle.

An in-ice interferometric phased array coherently 
combines signals from multiple low-gain antennas deployed down 
sub-surface boreholes with proper time delays 
to account for distances between antennas to effectively increase the gain of the system of antennas for incoming plane
waves from a given direction. Many different sets of delays of signals from the same antennas can create multiple
effective antenna beam patterns that would together cover the same solid angle as each
individual antenna but with much higher gain.  The closer the antennas
are physically, the fewer beams are needed to cover a given solid angle.  

This paper addresses the assumption that a phased array made of closely packed antennas receives uncorrelated noise
in each antenna. 
We show using realistic detector designs in both an anechoic chamber and in the ice in Greenland that thermal noise is 
uncorrelated between antennas. 
In developing phased arrays for use in the lab, we also demonstrate how the beamforming technique 
can be used for impulsive signals in practice.

In Section~\ref{sec:thermalNoise},
we discuss measurements of thermal noise correlation between closely-spaced antennas, relevant for 
an interferometric phased array trigger.  Section \ref{sec:beam} details a validation of the 
beamforming technique
for impulsive signals in an anechoic chamber.  
In Section~\ref{sec:trigger}, we discuss the implications of beamforming for 
a realistic triggering scheme.
Section~\ref{sec:site} reviews and details new measurements of the 
relevant ice and noise characteristics of Summit Station, Greenland,
the site where we performed an {\it in situ} noise correlation studies of a prototype detector.  
We conclude in Section~\ref{sec:conclusions}.

\section{Thermal Noise Correlation Studies}
\label{sec:thermalNoise}
One of the underlying assumptions in the interferometric phased array calculations is that the 
thermal noise measured by each antenna in the array is uncorrelated with the thermal noise measured in its
nearest neighboring antenna.  The level at which thermal noise signals are correlated between antenna channels
is one factor that determines the effective gain achieved by phasing together many antennas.  In the limit
of fully overlapping antennas, the thermal noise observed from the ice ($\sim$250~K) would be completely
correlated, and the noise from the system would be completely uncorrelated ($\sim$ 75~K
for the systems described in this paper).  
To determine how closely packed the antennas in a phased array can be without
introducing a significant correlated noise contribution, we
performed tests in an anechoic chamber and designed a simulation of thermal noise to compare to the measurements.

\subsection{Noise Correlation Measurements in an Anechoic Chamber}
\label{sec:ncmeas}

\subsubsection{Measurement Setup}
\label{sec:setup}
\begin{figure*}[!htbp]
      \begin{center}
        \includegraphics[width=16cm]{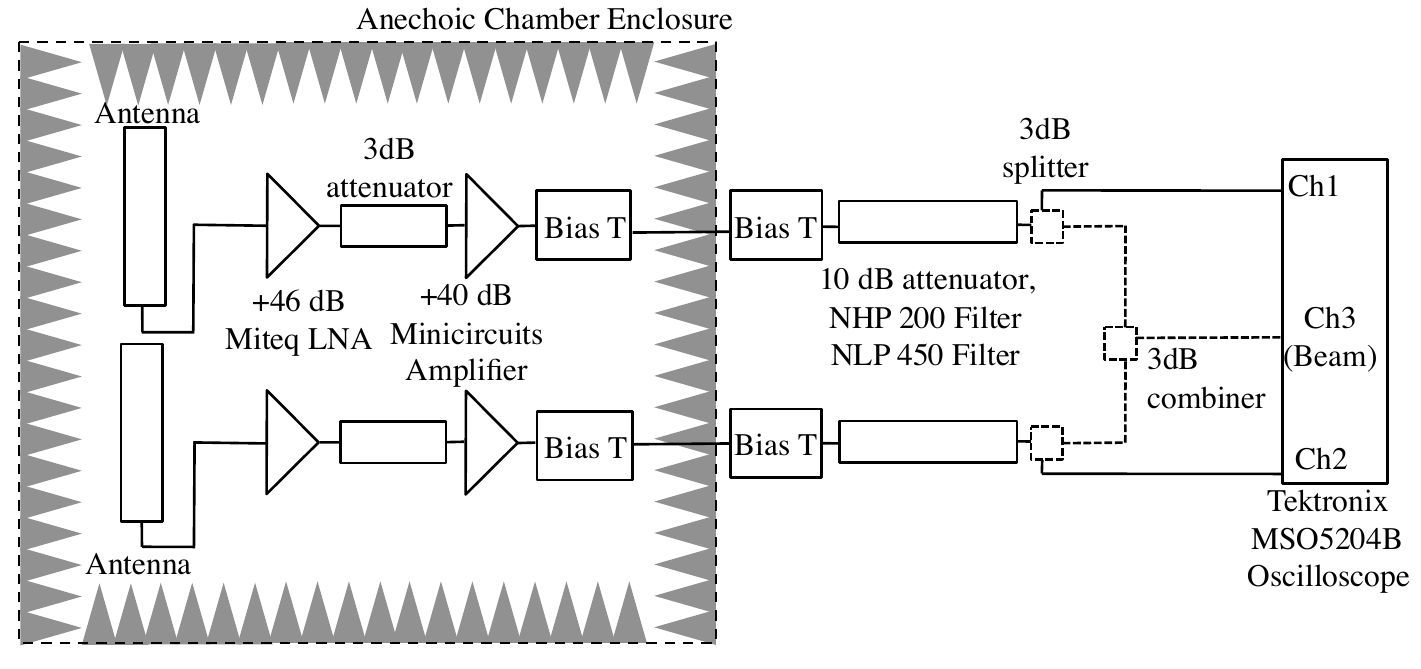}
    \end{center}
    \caption{A schematic of the setup in the anechoic chamber for thermal noise correlation testing and 
      validation of beamforming.  For thermal noise correlation testing, there were no splitters or combiners
      before the oscilloscope.  For validation of beamforming, described in Section~\ref{sec:beam}, 
      we added 3~dB splitters to each antenna
      channel and combined the signals to form a beam in hardware (shown with the dashed lines).  We 
      also set up a transmitter 4~m away inside the chamber for the measurements described in Section~\ref{sec:beam}, which is shown schematically in Figure~\ref{fig:antennaLayout}.} 
    \label{fig:anechoicSchematic}
  \end{figure*}

We performed noise correlation measurements using a simple system in an anechoic chamber.
Figure~\ref{fig:anechoicSchematic} shows a schematic diagram of the system setup, which consists of two 
antennas laid out end-to-end, with each antenna in its neighbor's null.  
Signals from each antenna were amplified by a dual-stage front-end amplifier chain that included a 46~dB 
low-noise amplifier (MITEQ AFS4-00100200-10-15P-4) and a 40~dB amplifier (Mini-Circuits ZKL-1R5) separated 
by a 3~dB attenuator.  DC power for the amplifiers was
carried through the radio frequency (RF) 
cable, coupled by bias tees inside and outside the anechoic chamber. Signals were then 
filtered using a Mini-Circuits NHP-200 and NLP-450 or NLP-600, depending on the type of antenna used for 
the test.  For all antenna types, which we discuss in Section~\ref{sec:antennas}, we used the NLP-600, 
except for the broadband dipole antennas that we developed, where
we used an NLP-450.  We used Times Microwave LMR-240 and LMR-400 cable, and cable lengths were identical in each 
signal chain.  The noise temperature of each channel was $\sim75$~K, dominated by noise from the 
front-end amplifier.  
Signals were then read out using a Tektronix MSO5204B oscilloscope, sampling at 5~GSa/sec.  
The walls of the anechoic chamber were between 1~m and 3~m from the antennas.

We changed the spacing between the antennas, ranging from as close as physically possible to a distance of over
1.5~m between antenna feeds, to measure the level of correlated noise between channels as a function
of the distance between the antennas. 

\subsubsection{Antennas Used}
\label{sec:antennas}

We used five different pairs of antennas for measurements in the anechoic chamber.  
We used two types of commercial antennas:
folded dipole antennas from Telewave, Inc. (ANT275D and ANT400D)
with bandpasses at 230-330~MHz and 360-450~MHz 
respectively\footnote{http://www.telewave.com/products/antennas/pdfs/TWDS-7048.pdf}
\footnote{http://www.telewave.com/products/antennas/pdfs/TWDS-7079.pdf}.  We also took data with two
types of antennas that have been developed for the ARA experiment.  
The first is a broadband bicone antenna, used by ARA to detect vertically-polarized
signals, and the second is a 
slot antenna that is sensitive to horizontal polarization when the antennas are deployed 
in boreholes~\cite{araWhitepaper}.  The bicone antennas have a bandpass of 150-850~MHz and the slot antennas
span 200-850~MHz.  

\begin{figure*}[!htbp]
      \begin{center}
        \includegraphics[width=15cm]{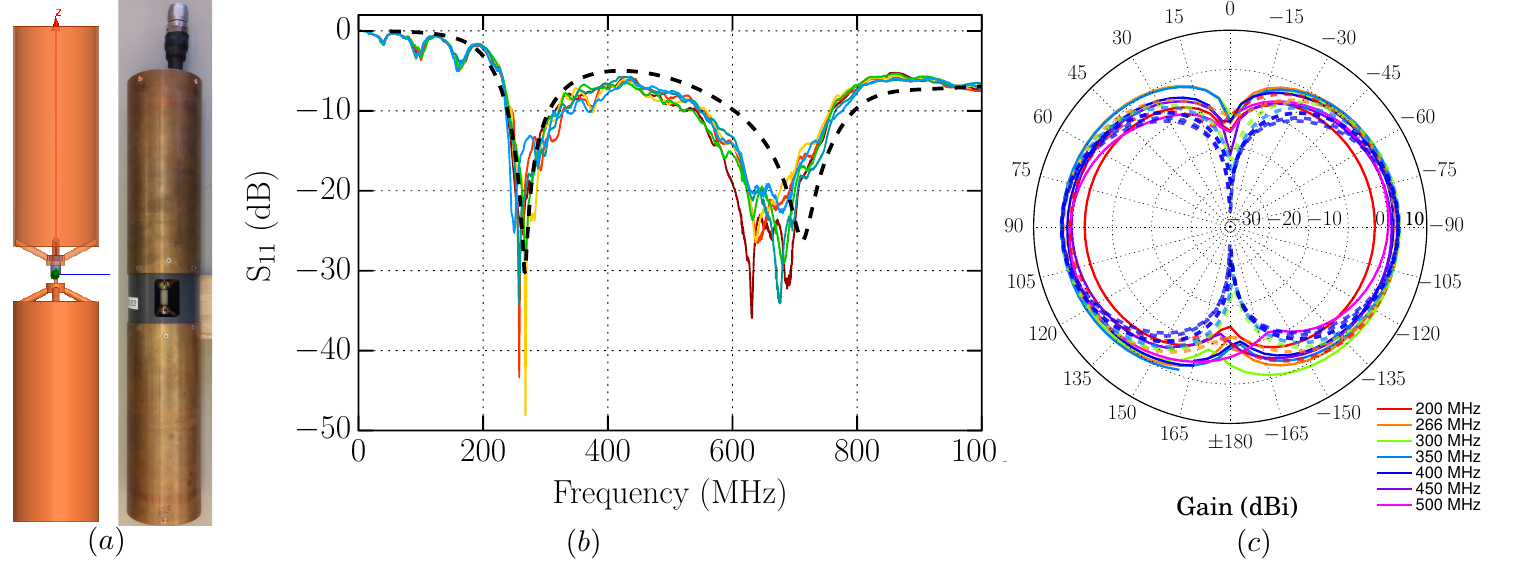}
    \end{center}
    \caption{(a) A simulated (using HFSS, left) and constructed (right) broadband dipole. (b) 
      The reflective S-parameter ($\mathrm{S}_{\mathrm{11}}$) for all the constructed antennas 
      (colored, solid lines) in air compared with the HFSS simulation (dashed).  These antennas have
      good transmission response from 230~MHz to 800~MHz, demonstrated by the small ($<-6$~dB)
      $\mathrm{S}_{\mathrm{11}}$ value over that frequency range.
      (c) Radiation patterns (solid lines) compared with HFSS simulations (dashed lines).} 
    \label{fig:antennas}
  \end{figure*}

Additionally, we took data with 
broadband dipole antennas that we developed, which we describe in detail here.
Each antenna consists of two 20~cm copper cylinders, 
each with a diameter of 8~cm. The two sides of the antenna are connected at the feed with 0.64 cm copper rods. 
The antennas are read out via a 1.3~cm diameter Heliax cable. A 5.1~cm polyvinyl chloride 
collar separates the two halves of the antenna and provides strain relief at the feed.  An antenna and the 
simulated design are shown in Figure~\ref{fig:antennas}a. 

The antenna frequency range is chosen to match
the expected Askaryan emission and the physical constraints of in-ice deployment
in boreholes. 
The Askaryan signal is coherent up to a few~GHz, but undergoes less 
propagation loss at low frequencies in glacial ice, 
so we preferred antennas in the 100-600 MHz range. 
The diameter of the dipoles was chosen to increase the bandwidth while ensuring 
that they would fit within reasonable sub-surface in-ice boreholes.

The antenna response, represented by the reflective S-parameter ($\mathrm{S}_{\mathrm{11}}$)
in Figure~\ref{fig:antennas}b, reaches 
its first resonance at 262 MHz in air, which is reduced to 196 MHz when surrounded by glacial ice ($n=1.78$). 
A simulation of the antenna response, constructed 
using HFSS\footnote{http://www.ansys.com/Products/Electronics/ANSYS-HFSS}, 
is shown with a dashed line in Figure~\ref{fig:antennas}b.
The higher-frequency dip corresponds to a second mode of the antenna. Simulations of the beam pattern 
in the antenna's E-plane  
indicate that the beam pattern becomes slightly more directive with increasing frequency,
rather than forming an additional null. However, the measured E-plane radiation pattern is no longer azimuthally symmetric above 450 MHz. 
The boresight gain is within 3~dB of its maximum of 2.9~dBi between 220~MHz and 755~MHz.

\subsection{Thermal Noise Simulation}
We model thermal noise in a multiple antenna system as the sum of multiple noisy transmitters following Reference~\cite{goodman}. Consider antennas $A$ and $B$ located distances $r_{An}$ and $r_{Bn}$ from a transmitter $n$ with random uncorrelated phases but equal amplitudes $\psi_0$. The statistical average of the cross-correlation coefficient, $\langle \mathbf{C}_{AB} \rangle$, of many sources is non-zero only when considering terms from a single transmitter, giving an average cross-correlation coefficient of
\begin{equation}
\langle \mathbf{C}_{AB} \rangle = \psi^2_{0} \Bigg \langle \sum_{n=1}^{N}  \frac{e^{-i k (r_{An}- r_{Bn} )}}{r_{An} r_{Bn}} \Bigg \rangle .
\end{equation}

Two cases may result in correlation between the antennas. If the transmitters are arranged in an arbitrary configuration, but there are boundaries, such as the ice-sky horizon that is relevant for balloon-borne neutrino experiments, partial correlation exists near the bounds. When the antenna separation and distance between the transmitters are much smaller than a wavelength, the source is unresolved and the average correlation is non-zero, as may be found with a closely packed phased array deployed down a borehole. Generally speaking, the correlation of the noise between a pair of antennas depends on a combination of the antenna separation and spatial distribution of the transmitters.

For comparison to the anechoic chamber measurements, we treat sources of thermal noise as oscillators 
each with an amplitude drawn from a 
Rayleigh distribution corresponding to a temperature of 300~K 
and a random phase drawn from a uniform distribution.  
We create $10^5$ of these oscillators, each with a 
frequency drawn from a uniform distribution over a 2~GHz bandwidth.  
The oscillators are thrown uniformly on a spherical shell that has a radius of 5~m and is centered 
on the point between two simulated antennas.
We weight the amplitude of each oscillator
by the sine of the incident angle on the antenna feed, where a zenith angle of zero corresponds to the location of
the null of the antenna, to approximate a dipole antenna beam pattern.  
We add the electric fields, weighted by the sine of the phase at each antenna, 
for all generated oscillators in Fourier space for each antenna and filter the result using 
the magnitude of the 
frequency response of the filters we used in thermal noise correlation measurements (see Section~\ref{sec:ncmeas}).  
Because the phase of each oscillator is chosen randomly from a uniform distribution, we do not need to 
take into account the phase response of the filters, since the resulting phase 
would still be randomly drawn from a uniform distribution. 
Uncorrelated noise is added to the simulated thermal noise over the same bandwidth for each channel
that corresponds to the 75~K system temperature for each channel by
generating noise with a spectral amplitude drawn from a Rayleigh distribution
and a random phase drawn from a uniform distribution over the same bandwidth as the thermal noise.
We generate simulated noise traces at each antenna by taking the inverse Fourier transform of
the filtered spectra. 
By generating many sets of noise traces for antennas with different feed-to-feed spacings, we can make
direct comparisons with measurements.  

We developed two independent simulations, which we used as a validation technique, 
and the results of the simulations are consistent with each other.

\subsection{Results}
\begin{figure*}[!htbp]
      \begin{center}
        \includegraphics[width=14cm]{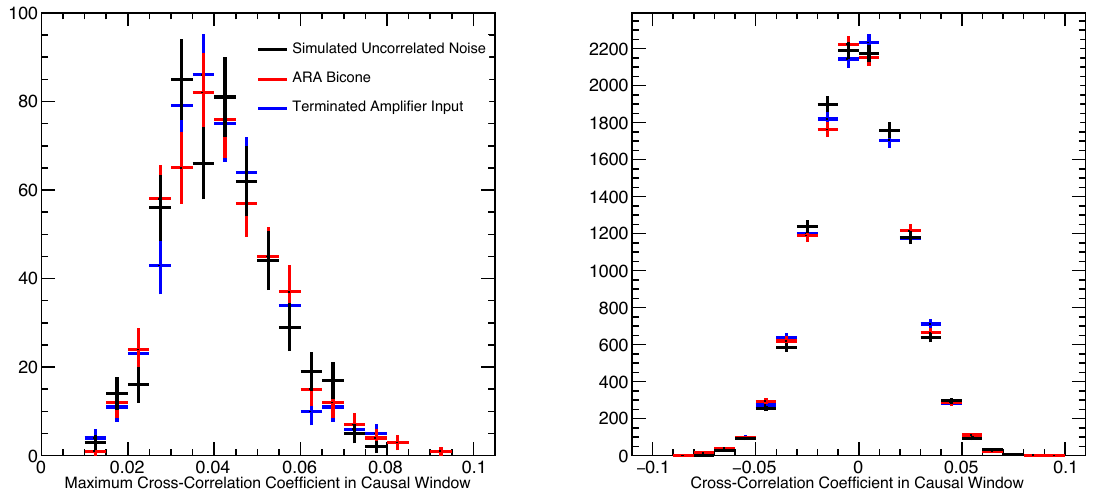}
    \end{center}
    \caption{Left: a histogram of cross-correlation coefficient 
      values between the two antenna channels over a time window
    $\pm2.4$~ns around zero time delay for each of $\sim500$ recorded events compared to results from the simulation.  
    This time window corresponds to possible time differences between
    antennas for incident plane wave signals from an arbitrary direction, determined by the 
    spacing between the antenna feeds.  
    Right: a histogram of the maximum of the absolute value of cross-correlation
    coefficient values between the two antenna channels in the causal time window for each of $\sim500$ recorded events.  
    The correlation values for a data set with the inputs to the front-end amplifiers terminated with a 50~$\Omega$
    load are shown in blue, and the values for the configuration with the ARA bicone antennas with their feeds 
    separated by 0.73~m (the closest possible physical spacing) are shown in red.  
    The black line shows results from
    the thermal noise simulation for uncorrelated noise inputs.  $\sqrt{N}$ error bars are shown per bin. } 
    \label{fig:noiseCorrelation}
  \end{figure*}

We run identical analyses on simulated noise data sets 
and on noise correlation data taken in the anechoic chamber.
Figure~\ref{fig:noiseCorrelation} shows the results of the noise correlation 
measurements for the ARA bicone antennas spaced as closely as physically possible (a feed-to-feed distance of 0.73~m), 
compared to measurements with the inputs to the front-end amplifiers terminated with a 50~$\Omega$ load and
results from the thermal noise simulation.  
The configuration with terminated amplifier inputs 
is used as a measure of the baseline level of correlated noise in the system between
the two antenna channels.

The left-hand panel of Figure~\ref{fig:noiseCorrelation} 
is a histogram showing the cross-correlation coefficient between
the two antenna channels over a causal time window ($\pm$2.4~ns) for each of $\sim500$ recorded events. 
This causal time window corresponds to possible time differences between
antennas for incident plane wave signals from an arbitrary direction, determined by the 
spacing between the antenna feeds. The causal time window ($\Delta t$) is determined by
$\Delta t = \Delta d / c $, where $\Delta d$ is the
feed-to-feed spacing between the antennas (0.73~m), and $c$ is the speed of light.
We define the cross-correlation coefficient, $\mathbf{C}$, between two time-domain waveforms, 
$x(t)$ and $y(t)$, at a relative time delay $\tau$ between the waveforms, as

\begin{equation}
\mathbf{C}(x(t),y(t), \tau) =  \frac{1}{N \sigma_x \sigma_y} \sum_{i=1}^{N}  (x(t_i) - \bar{x})  (y(t_i+\tau)-\bar{y}),
\end{equation}
where $N$ is the number of points in each waveform, $\sigma_x$ and $\sigma_y$ are the 
standard deviation of each waveform, and $\bar{x}$ and $\bar{y}$ are the mean of each waveform.
Since the causal time window ($\pm$2.4~ns) is much smaller than the length of each trace (2000~ns), 
the bias of the cross-correlation across the causal window is small ($\sim0.2$\%).

The right-hand panel of Figure~\ref{fig:noiseCorrelation} 
is a histogram of the maximum of the absolute value of cross-correlation coefficient values shown 
in the left-hand plot for each recorded event.  The 
three overlaid histograms are consistent within statistical errors, shown on the histograms.
There is no significant correlation seen between the two channels
when compared to the data taken with terminated amplifier inputs,
even when the antennas were spaced as closely as physically possible.  The correlation observed is 
consistent with simulated uncorrelated noise, shown by the black line in
Figure~\ref{fig:noiseCorrelation}.  

\begin{figure}[h!]
      \begin{center}
        \includegraphics[width=8cm]{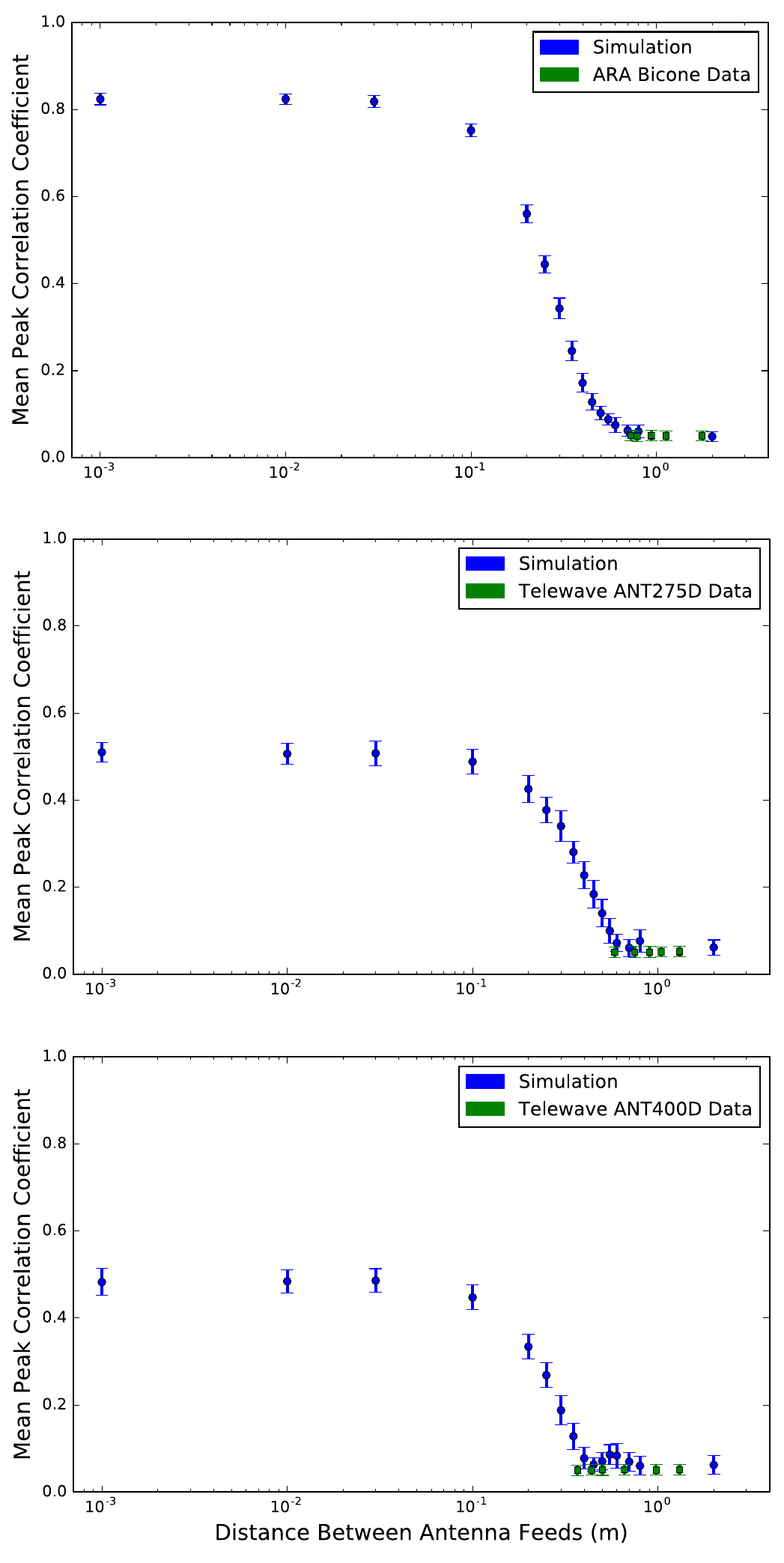}
    \end{center}
    \caption{The peak cross-correlation coefficient
      averaged over 500 simulated or measured events in a $\pm6.0$~ns time window, which
      is the causal window 
      for the largest feed-to-feed spacing for which we made measurements ($\sim2$~m).
      The error bars represent the standard deviation of the peak cross-correlation coefficient
      values across the 500 events.  The data for each antenna goes to the smallest physically-allowed
      spacing.  
      Predictions from the simulation are shown in blue and results from the measurements are shown in green. Top: 
      ARA bicone antennas. Middle: Telewave ANT275D antennas.  Bottom: Telewave ANT400D antennas.} 
    \label{fig:noiseSimulation}
  \end{figure}

We repeated this measurement with the antennas 
spaced farther apart and with a variety of types of antennas (Telewave ANT275D, Telewave ANT400D, and the 
ARA bicone antennas), and all results are consistent with no observed correlation between adjacent antennas.  
Figure~\ref{fig:noiseSimulation} shows the results of a comparison between the thermal noise 
simulation and data taken 
with each of the types of antennas at a variety of feed-to-feed distances.  For these antennas,
the feeds are at the centers of the antennas, and a spacing of zero in the simulation corresponds to 
fully overlapping antennas.  Shown is the peak cross-correlation coefficient
averaged over 500 simulated or measured events in a $\pm6.0$~ns time window, which is the causal window 
for the largest feed-to-feed spacing for which we made measurements ($\sim2$~m).
We maintain a consistent time window in analysis and simulation at each distance 
to keep the trials factor the same across all distances shown.
The error bars represent the standard deviation of the peak cross-correlation coefficient
values across the 500 events.  

At very small hypothetical feed-to-feed distances, 
the simulation approaches a maximum correlation coefficient that corresponds to the fraction of 
total noise in each channel due to the 300~K thermal noise compared to the total noise (the sum of the thermal noise and the 75~K system temperature). The thermal noise is filtered by both the antenna and the in-line filters, whereas the 75~K system temperature filtered by the in-line filters only, which are broader band than the Telewave antennas. The 75~K system temperature is independent between channels.

At large spacings, the simulated data has a peak 
correlation coefficient of 0.05 for all three antenna types, which would change with 
different allowed time windows.
True physical noise correlation, beyond the correlation level expected from the noise 
statistics alone (in our case $\sim5$\%), 
is not seen in simulation until the antennas are closer together than physically possible.  

\begin{figure}[h]
      \begin{center}
        \includegraphics[width=8cm]{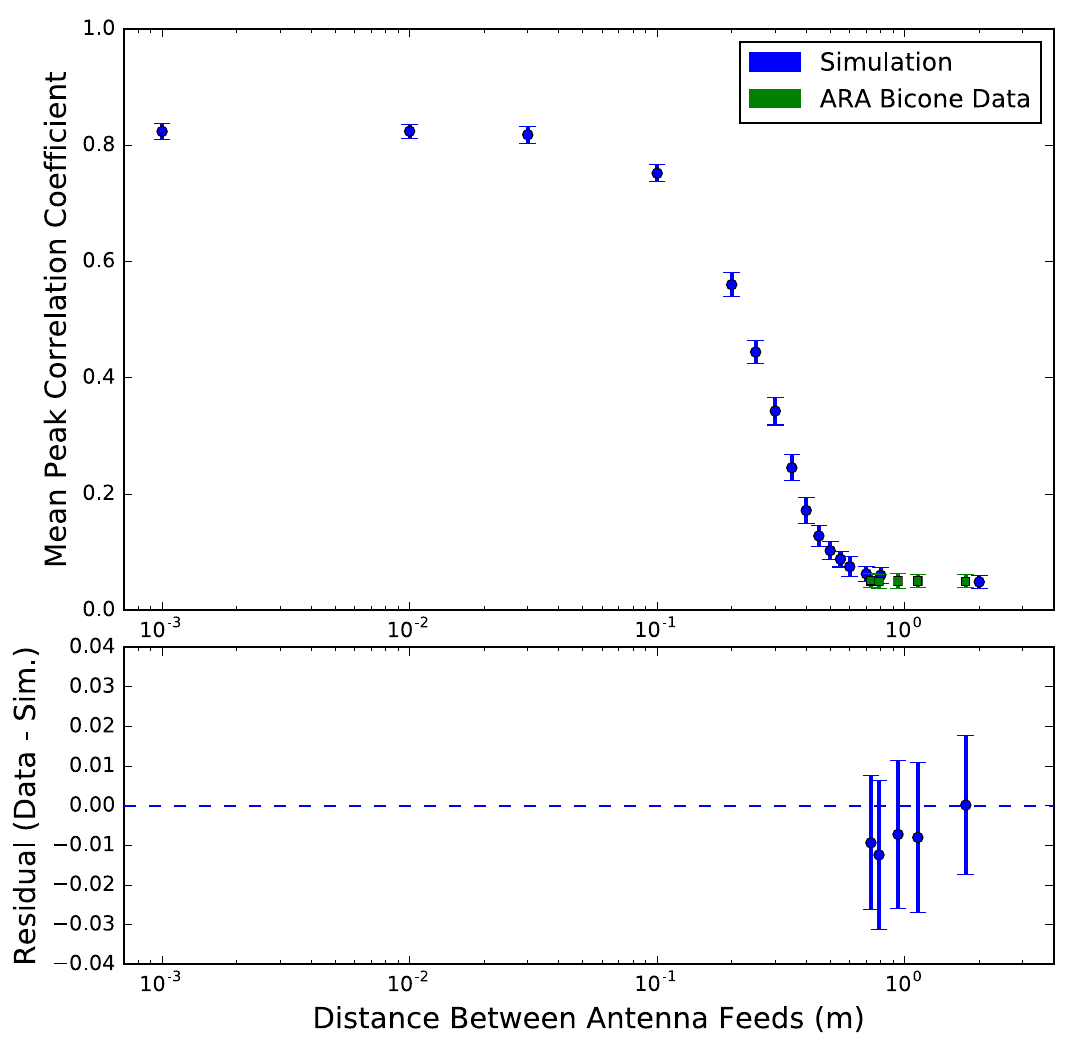}
    \end{center}
    \caption{Top: The peak cross-correlation coefficient for the ARA bicone antennas.  Predictions from the 
simulation are shown in blue and results from the measurements are shown in green. (The same as the top panel in Figure~\ref{fig:noiseSimulation}.)  Bottom: The difference between the measured and simulated values for the ARA bicone antennas.  The error bars represent 
the quadrature sum of the error bars shown in the top panel.} 
    \label{fig:residuals}
  \end{figure}

At physically allowed spacings where measurements are possible,
({\it e.g.}, $>0.73$~m for the ARA bicone antennas), the measured and 
simulated noise correlation for all antenna types agrees to within $<50$\% of the measured and 
simulated standard deviation. The difference between the measurements and the simulation for the ARA
Bicone antennas is shown in Figure~\ref{fig:residuals}.
We attribute these small differences to imperfect assumptions in the simulation, such as ignoring differences in antenna
phase response as a function of angle, which would serve to further decorrelate channels.  

\section{Validation of Beamforming Technique}
\label{sec:beam}

We also performed a simple validation measurement 
of the beamforming technique in the anechoic chamber.  We used a system configuration shown in the schematic in
Figure~\ref{fig:anechoicSchematic}. 
The configuration was identical to the system described in Section~\ref{sec:thermalNoise}, except for the following 
difference: after the second bias tee, we split the signal from each antenna channel and combined one branch of the 
signal from each channel to form a single beam that corresponds to zero time delay between the channels.  
This is shown with the dashed line in Figure~\ref{fig:anechoicSchematic}.
This beam is broadside to the antennas, at the highest-gain angle of each antenna.  For this measurement, 
we used the broadband dipole antennas that we developed (see Section~\ref{sec:antennas}).
  \begin{figure}[h]
      \begin{center}
        \includegraphics[width=7cm]{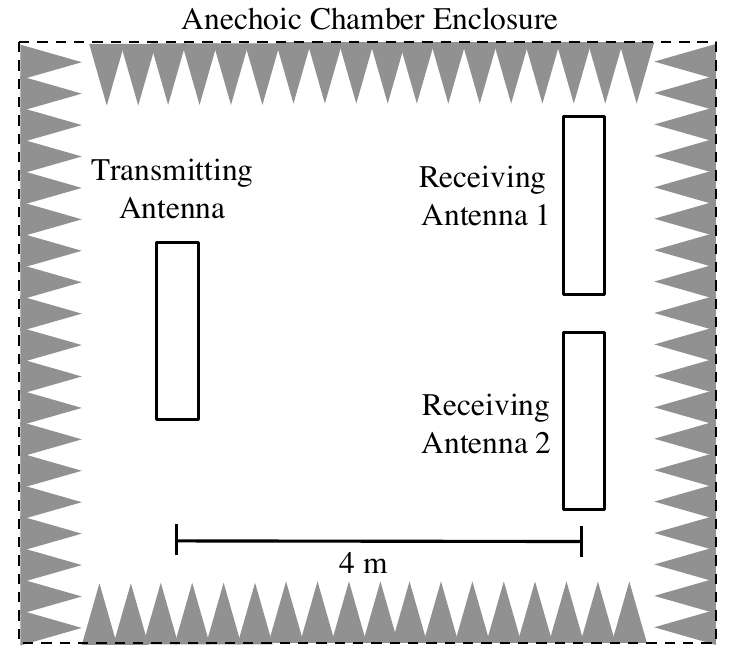}
    \end{center}
    \caption{A schematic of the layout of the antennas for the transmission measurement described in 
      Section~\ref{sec:beam}. 
      } 
    \label{fig:antennaLayout}
  \end{figure}

  \begin{figure}[h]
      \begin{center}
        \includegraphics[width=9cm]{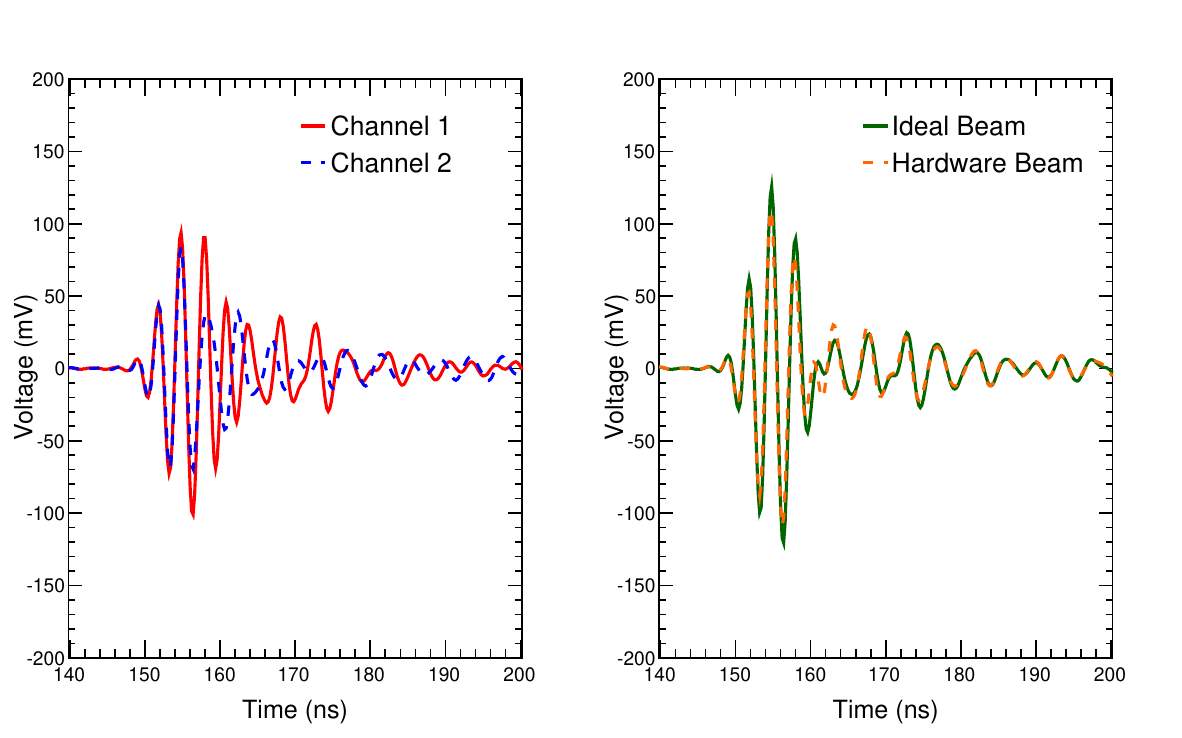}
    \end{center}
    \caption{Left: an overlay of waveforms, averaged over 500 events, for Channels 1 and 2 in the 
      system shown in Figure~\ref{fig:anechoicSchematic}, when transmitting
      a fast impulse to the antennas.  Right: an overlay of waveforms, averaged over 
      500 events, for the hardware-summed beam recorded in Channel 3 in the configuration shown in 
      Figure~\ref{fig:anechoicSchematic} and the coherent waveform calculated directly from the averaged
      waveforms shown in the left-hand panel.  This test uses the broadband 
      dipole antennas described in~\ref{sec:antennas}. 
      } 
    \label{fig:beamAnechoic}
  \end{figure}

We sent broadband, impulsive signals to the two receiving antennas from an identical transmitting antenna
that was 4~m away inside the anechoic chamber, and was positioned broadside to the receivers for maximum
transmission strength in the direction of the formed beam.  A layout of the antennas inside the chamber
for this transmission measurement is shown in Figure~\ref{fig:antennaLayout}. 
The signals are generated using an Avtech AVP-AV-1S-C-P pulse generator, and are filtered
before transmission with a Mini-Circuits NHP-200 and an NLP-450 filter.  

Figure~\ref{fig:beamAnechoic} shows the results of the measurement.  The left-hand panel shows the impulsive
signal received in the two independent antenna channels, averaged over 500 events recorded on the oscilloscope.
Each channel had a slightly different impulse response, dominated by differences between individual antennas.  Each
channel also had a slightly different gain and noise temperature, dominated by differences in the individual 
front-end low-noise amplifiers. 
The right-hand panel shows the beam that we formed in hardware, compared to the ideal beam that should be formed, 
calculated by summing the waveforms shown in the left-hand panel.  
Some difference is evident between the beam formed in hardware and the ideal version 
formed in analysis; the
peak-to-peak voltages agree to within 15\%, and the difference is consistent with the additional loss
expected from the splitters and combiners added to the system to create the hardware-formed beam.

Before averaging together many waveforms to produce the high signal-to-noise ratio (SNR) 
waveform shown in Figure~\ref{fig:beamAnechoic},
we calculate the SNR of the signal that was received in each channel, including the hardware-formed beam,
compared to the noise level in the system.  We define SNR as half of the signal's peak-to-peak voltage divided by
the RMS of the noise ($\frac{V_{\mathrm{pk2pk}}}{2 \: \sigma}$). 
The results are shown in Table~\ref{tab:beam}.  We then calculate
the ideal beam that should be formed and the noise level corresponding to superimposed noise from the 
individual antenna channels.  Note that because of the differences between the two signal chains, 
we do not expect to see the $\sqrt{2}$ improvement in SNR that should be obtained for identical
antennas and signal chains.  Instead, we expect to see the improvement shown in the last line of Table~\ref{tab:beam}.
The hardware beam matches the ideal beam well in terms of SNR, although 
both the signal level and noise level suffer slightly from additional losses introduced by the additional
components in the signal chain at the 10\% level.

\begin{table}
\begin{center}
\begin{tabular}[c]{|l|c|c|c|}
\hline 
& Noise  & Signal  & SNR \\
&$V_{\mathrm{RMS}}$ & $V_{\mathrm{pk2pk}}$ & ($\sigma$)\\
& (mV) & (mV) &  \\
\hline
Antenna Ch. 1 (measured) & 37.6 & 194.7 & 2.6 \\
Antenna Ch. 2 (measured) & 28.5 & 154.3  & 2.7  \\
HW Beam (measured) & 35.5 & 217.2 & 3.1  \\
Ideal Beam (calculated) & 37.2 & 246.1 & 3.3 \\
\hline
\end{tabular}
\end{center}
\caption[]{\label{tab:beam} A summary of the validation of the beamforming technique using data taken
  in an anechoic chamber.  Shown are the noise levels, signal amplitude, and SNR of the signals measured 
  for each antenna channel and the beam formed in hardware, as well as the properties of the 
  expected ideal beam calculated using data from each antenna channel.
 }
\end{table} 

\section{Trigger Studies for  Impulsive Events}
\label{sec:trigger}
We investigate the trigger rate and efficiency of a broadband phased array using the
anechoic chamber measurements in combination with a simulation study of both noise and transient signals. 
A linear array of three Telewave ANT400D folded dipole antennas was placed in the anechoic chamber
approximately 4~m from an ARA bicone antenna that was used as a transmitter (see Section~\ref{sec:antennas}). 
The Telewave antennas 
are mounted on a steel ground mast with a separation of 36~cm between the antenna phase centers and the ground mast. 
The antennas are spaced at a pitch of 55~cm. Each receiving antenna has the same signal chain as shown in 
Figure~\ref{fig:anechoicSchematic}, and the signals are recorded at 5~GSa/sec.

We model the digital phasing and event triggering of a linear antenna array. 
The antenna channels are formed into beams by digitally delaying and coherently summing the 
individual antenna channels. 
A relative delay is applied between the digitized waveforms received at a pair of antennas, which corresponds
to a beam angle, $\theta_n$, of

\begin{equation}
n \: \Delta t = \frac{d}{c} \sin \theta_n ,
\label{eqn:delay_beam}
\end{equation}
where $n$ is an integer, $\Delta t$ is the the sampling time interval, $c$ is the speed of light in the medium, 
and $d$ is the baseline between the pair of antennas. Many beams can be formed simultaneously in hardware 
to cover a wide angular range, either digitally using FPGA or 
analog ({\it e.g.}, as described in Section~\ref{sec:system}). For this trigger study, we form beams
in analysis by coherently adding waveforms from individual antenna channels.
For a fixed antenna spacing, the sampling rate sets
the number of independent beams and the granularity of the angular coverage.

With an antenna spacing of 1~m and a sampling rate of 2~GSa/sec, we use Equation~\ref{eqn:delay_beam} 
to find that 17 independent beams, formed using the smallest 1~m baseline, are needed to cover the 
elevation angle range between -$45^\circ$ to +$45^\circ$. The beamwidth is determined by the number of 
antennas in the array. As the beamwidth narrows, more beams can be added to fill the coverage 
gaps by utilizing correlations between antennas at all baselines in the array.

The optimal sampling interval for the power calculation is related to the dispersion of the signal going into the 
trigger. For the case in which the impulse response of the system is deconvolved before triggering, it may be 
optimal to sample the power at every data sample because the signal power is contained in a short time interval. 
However, the system response is not unfolded from the signal in the baseline algorithm 
described here. A more efficient algorithm for the dispersive pulses recorded in the anechoic chamber 
calculates the power every eight samples, thereby including the majority of the power from a single transient pulse. 

To detect transient events, a time-windowed power calculation is performed
on each coherently summed beam, given by

\begin{equation}
P_{window} = \frac{1}{N_{s} \: Z_{L}}\sum_{j=1}^{N_{s}} \left(\sum_{i=1}^{N_{ant}} V_{ij} \right)^2,
\label{eqn:beampower}
\end{equation}
where $N_{s}$ is the number of digitized samples in the time window, $Z_{L}$ is the load impedance,
$N_{ant}$ is the number of antennas used in the beam, and $V_{ij}$ is the digitized voltage at sample $j$ within
the time window for antenna $i$.  

\begin{figure}[t]
      \begin{center}
        \includegraphics[width=9cm]{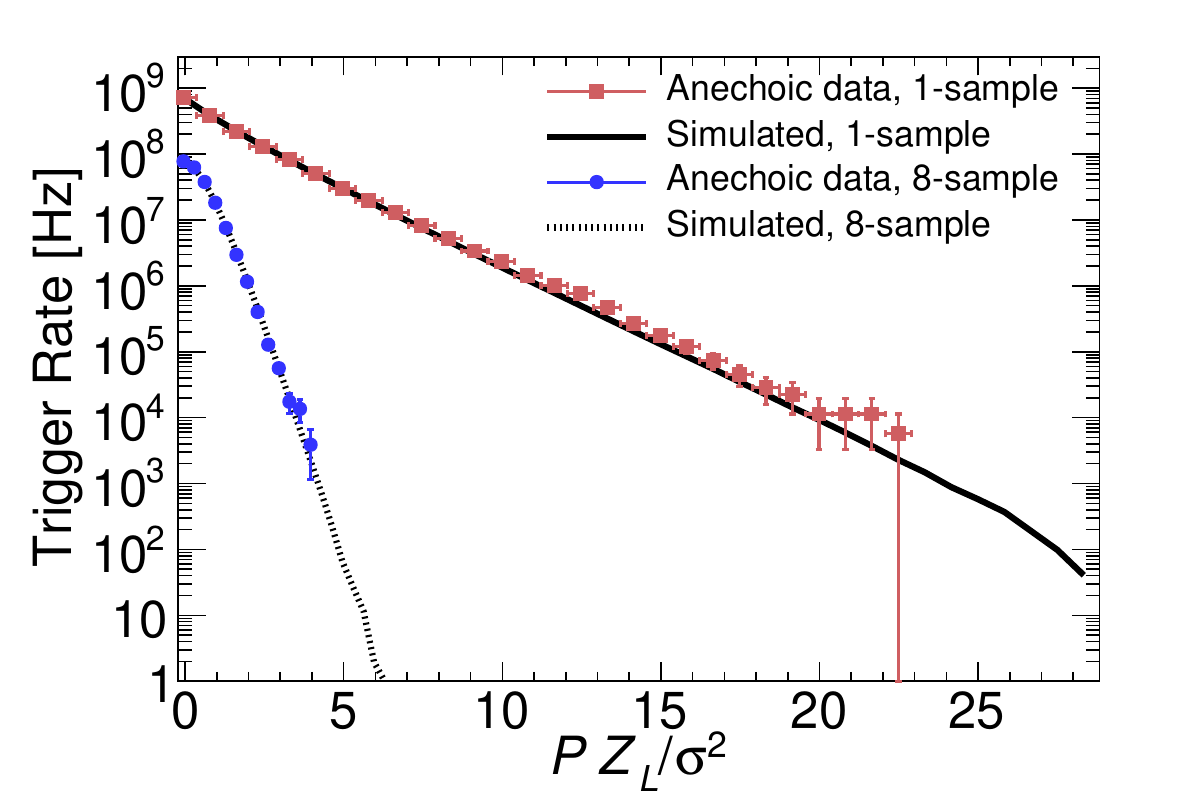}
      \end{center}
    \vspace{-20pt}
    \caption{Trigger rate vs. normalized power threshold, $P Z_{L} / \sigma^{2}$, for noise traces 
      taken in the anechoic chamber using the three Telewave antennas
      (data points) compared to simulations of thermal noise (solid lines).  
      Two trigger configurations are shown: 
      the blue data points indicate the trigger rate when the average power is taken over a 9.6~ns window and
      incremented every eight samples (4.8~ns). The red data points show the singles rate when the power is calculated 
      at every 
      sample point. We use the simulation to predict the threshold for rates lower 
      than several kHz due to the limited amount of data taken in the anechoic chamber.} 
    \label{fig:rates}
  \end{figure}

The relationship between the per-beam singles rate and trigger 
threshold is determined by $N_{s}$, as shown in Figure~\ref{fig:rates},
where the power threshold, $P$,
is normalized by the square of the noise RMS ($\sigma^2$) divided by $Z_{L}$. 
For these measurements, the data are down-sampled from 5~GSa/sec to 1.67~GSa/sec to reflect a 
Nyquist sampling rate for a system with a bandwidth of $\sim$800~MHz.

Two power calculations are shown in Figure~\ref{fig:rates}:
one in which the average power is calculated in a window of $N_{s}$=16 samples (9.6~ns) and sampled
every $N_{s}/2$=8~samples (4.8~ns), and a second in which the power in each sample is recalculated at every 
sample point.
With a power calculated from an 8-sample window, we find voltage thresholds, $\sqrt{P Z_{L}}$, of 2.34$\sigma$, 
2.21$\sigma$, and 2.07$\sigma$, for noise singles rates of 10, 100, and 1000~Hz, respectively, 
which corresponds to the range of achieved rates in radio-detection experiments.

  \begin{figure}[h]
      \begin{center}
        \includegraphics[width=8cm]{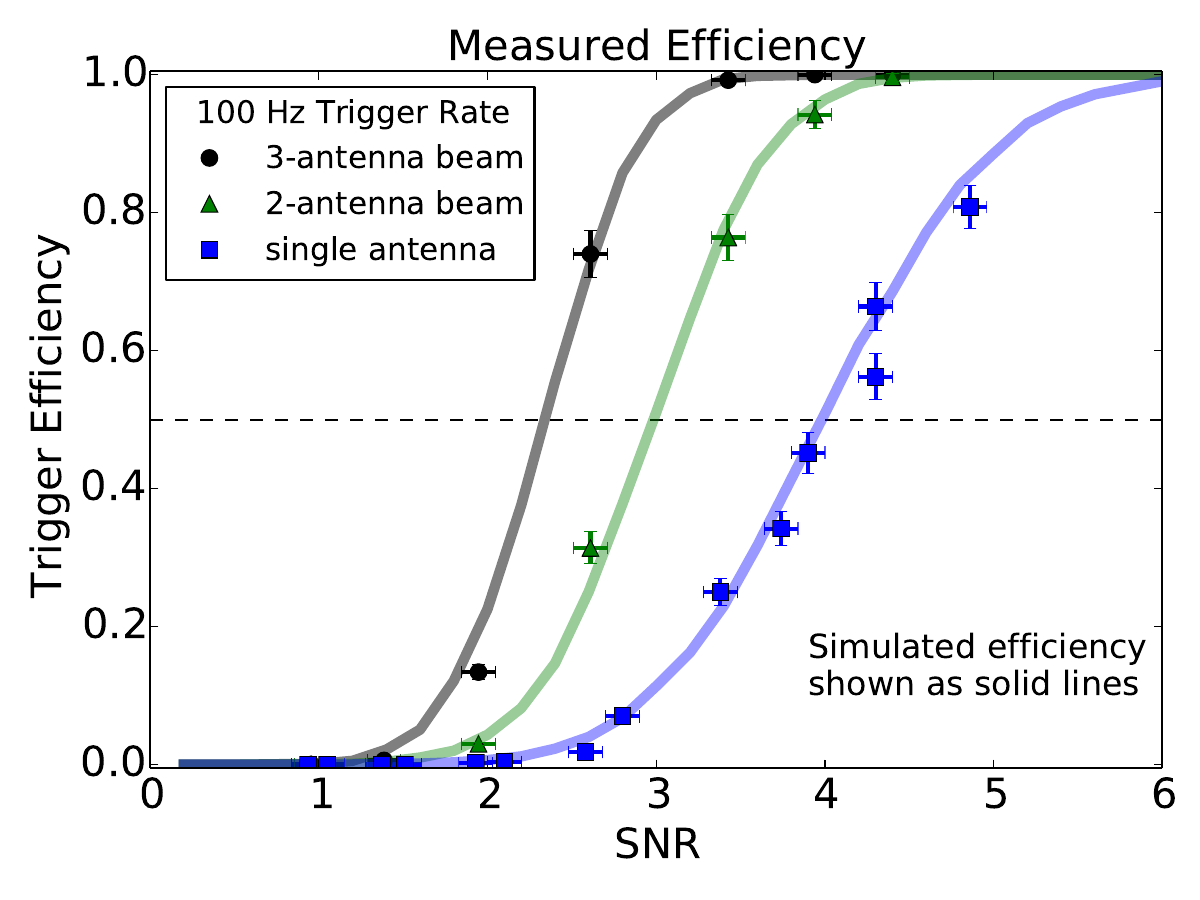}
        \includegraphics[width=8cm]{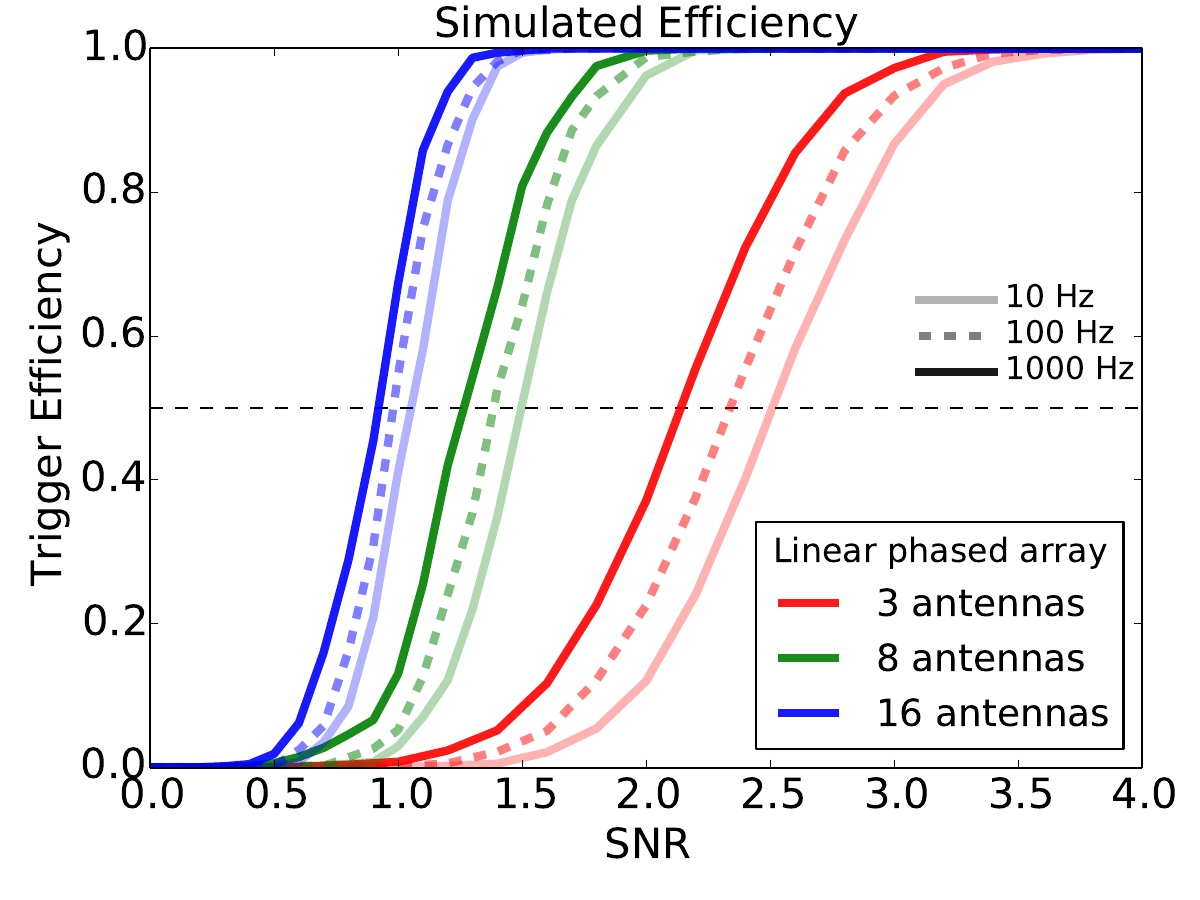}
    \vspace{-20pt}
    \end{center}
    \caption{Trigger efficiency vs. SNR. The top plot shows anechoic chamber results from a
      3-antenna array (data points) compared to a simulation of uncorrelated noise added to the
      system impulse response (lines). 
      The data points are taken from anechoic chamber measurements with a fast impulsive signal, and 
      the average power was calculated at 8-sample intervals using a power threshold corresponding to a 100~Hz 
      singles rate 
      (dashed line from~Figure~\ref{fig:rates}).  
      The bottom plot shows the simulated trigger efficiencies for 3, 8, 
      and 16-antenna broadband phased arrays in a single formed beam. 
      For each configuration, efficiency curves are drawn for per-beam trigger rates of 
      1~kHz, 100 Hz, and 10 Hz by the dark-solid, dashed, and light-solid line, respectively.} 
    \label{fig:efficiency}
  \end{figure}

To measure the trigger efficiency, event rates were measured from the 3-antenna array in the anechoic chamber 
while sending impulsive signals from 
the transmitting antenna at several attenuation levels.  For each attenuation setting,
we recorded 500 events. The SNR is defined as 
$\frac{V_{\mathrm{pk2pk}}}{2 \: \sigma}$ for a single antenna.  
We choose 100~Hz as a baseline per-beam trigger rate, comparable
to achieved trigger rates by currently-deployed radio experiments, such as ANITA.
We measure the efficiency for a single antenna, the 0$^\circ$ beam formed by using 2 antennas, and the full 
3-antenna 0$^\circ$ beam by comparing the power, calculated within the time window corresponding to the time 
when the pulse was transmitted using Equation~\ref{eqn:beampower},
to the appropriate threshold level for the chosen trigger rate. 
The measured trigger efficiency at a trigger
rate of 100~Hz, specified by Figure~\ref{fig:rates} using the 8-sample interval, is shown in the left plot in 
Figure~\ref{fig:efficiency}.  

These measurements compare well with simulation results shown by the solid lines in the left-hand plot of 
Figure~\ref{fig:efficiency}.  
The simulated curves include 5000 events in which the average impulse response of 
each antenna is added to the appropriate level of uncorrelated system noise (75~K for our system
plus 300~K of room temperature thermal noise).  We chose to simulate 5000 events to achieve standard
errors that are $<$1\%.

The simulation is extrapolated to larger phased arrays as shown on the right-hand 
plot in Figure~\ref{fig:efficiency}, and curves are shown for each size array 
at 10, 100, and 1000 Hz per-beam trigger rates, corresponding to the 
range of achieved rates in radio-detection experiments. 
The efficiency curves for the 8- and 16-antenna arrays
assume equal impulse response for each antenna channel. 
A 16-antenna linear phased array that is set to trigger 
on impulsive events in a single beam at 100 Hz achieves a 50$\%$ efficiency at a SNR of about one.   
This is an improvement of a factor of four in SNR over single-antenna trigger thresholds, which
reach 50\% efficiency at an SNR of about four at the same trigger rate.

\section{An In-Ice Phased Array at Summit Station}
\label{sec:site}
We have also studied thermal noise using an analog interferometric phased array in the ice at Summit Station in Greenland. In this Section, we describe the site, the analog phased array deployed there, and the noise environment observed with the phased array.

\subsection{Summit Station in Greenland}

Summit Station is a year-round NSF-operated site, located at N 72$^\circ$ 37' W 38$^\circ$ 28', 
near the highest point on the Greenland ice sheet.  Summit Station sits atop 3~km of
glacial ice, making it an ideal candidate site for radio detection of high energy neutrinos. 
At Summit Station, the density of the firn has reached 95\% that of glacial ice by a depth of
100~m~\cite{summitIceCore}.  At the South Pole, the density has reached 95\% that of glacial ice by a depth 
of 140~m~\cite{koci}.  Therefore, for a detector that is designed to sit below the bulk of the firn rather
than at or near the surface, Summit Station would require shallower drilling compared to the South Pole.



The radio clarity of the ice at a site sets the effective 
volume of ice in which neutrino interactions are observable 
for a given radio detector configuration, directly related to the sensitivity of the experiment.  
The depth-averaged field attenuation length at Summit Station has been measured to be 
$\langle L_\alpha \rangle =947^{+92}_{-85}$~m at 75~MHz~\cite{avva}.  To directly compare this measurement
with radio attenuation length measurements that have been made previously 
at other sites of developing and proposed neutrino detectors, 
such as the South Pole (the site of ARA) and Moore's Bay on the Ross Ice Shelf (the site of ARIANNA), 
this measurement 
is extrapolated to 300~MHz and averaged over only the upper 1500~m of ice, which is where the interaction vertex
would be for most neutrino events that are detectable by a surface or sub-surface detector.
Assuming the measured 
temperature profile of ice at Summit Station and measured dependence of attenuation length on frequency, 
this indicates an average field
attenuation length of $1022^{+230}_{-253}$~m over the upper 1500~m~\cite{avva}, compared to 
$1660^{+255}_{-120}$~m over the top 1500~m at 300~MHz at the South Pole~\cite{araWhitepaper} and
$411$~m with an experimental uncertainty of about 40~m
averaged over all depths for the 578~m thick Moore's Bay location~\cite{barrella,ariannaAtten}.

  \begin{figure}[h]
      \begin{center}
        \includegraphics[width=9cm]{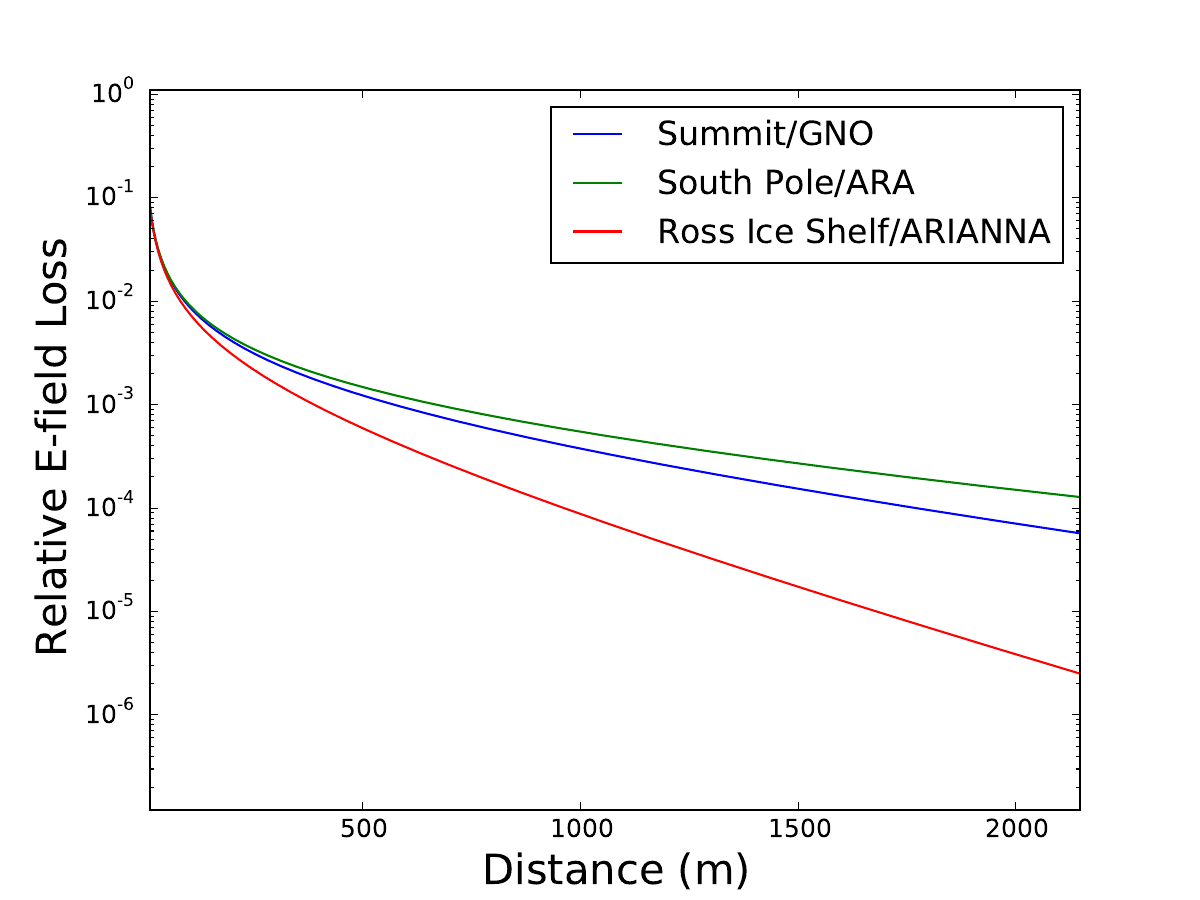}
    \end{center}
    \caption{The electric field loss at 300~MHz as a function of radio propagation distance through 
      ice at potential 
      sites for neutrino detectors (Moore's Bay, the South Pole, and Summit Station).} 
    \label{fig:eloss}
  \end{figure}

We can compare the electric field loss as a function of distance traveled through ice for
the three candidate sites, which is the metric that is directly related to neutrino effective volume for a 
given detector.
Figure~\ref{fig:eloss} shows the electric field loss as a function of radio propagation distance through 
ice at candidate sites for neutrino detectors (Moore's Bay, the South Pole, and
Summit Station), including both the $1/r$ geometric factor and the measured attenuation length at each site. 
For neutrino events, which typically occur hundreds of meters from the detector, the loss seen through the 
ice at Summit Station is comparable to the loss seen through the ice at South Pole, and is much less than at 
Moore's Bay.


\subsection{An In-Ice Phased Array at Summit Station}
\label{sec:system}

\begin{figure*}[!htbp]
      \begin{center}
        \includegraphics[width=10cm]{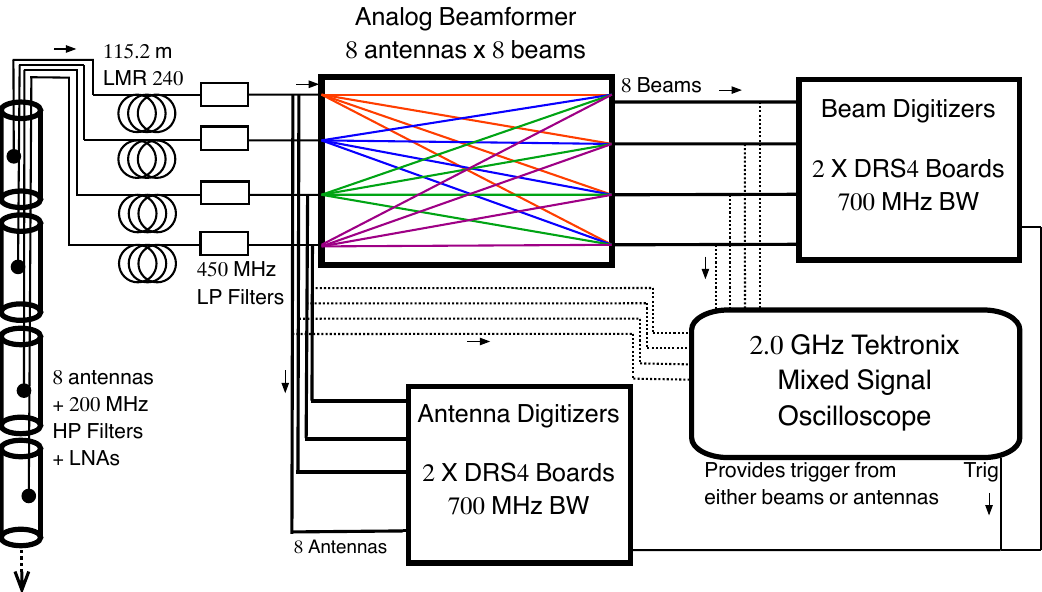}
    \end{center}
    \caption{A schematic of the system deployed in Greenland. Only four signal chains and 
      beam channels are shown for simplicity.} 
    \label{fig:systemDiagram}
  \end{figure*}

An instrument was deployed at Summit Station in June 2015 to characterize the site and validate
the phased array technique with 
an array deployed in the ice, as discussed in Reference~\cite{2015ICRCGNO}. 
The system used analog beamforming to combine signals from multiple antennas, 
as shown in the schematic of the RF signal chain in Figure~\ref{fig:systemDiagram}. 
The system used the same front-end amplifier system described in Section~\ref{sec:setup}.
Each front-end amplifier was preceded by a 200~MHz high pass filter to protect 
the first stage amplifier from the 8~MHz transmitter at Summit Station. 
Coaxial Times Microwave LMR-240 transmitted 
the antenna signal to the surface over 115~m of cable. Variations in the cable lengths led to sub-ns 
variation in arrival times between antenna channels. 

The signal from each 
antenna was split into two signal chain branches 
using 3~dB splitters. The first branch of the chain carried signals from 
individual antennas, while the second branch formed beams in the beamformer 
by splitting signals from each antenna eight ways,
propagating signals from each antenna through fixed delay lines of Times Microwave LMR-200, 
and combining signals into eight beams at fixed angles from the horizontal.
The LMR-200 delay lines ranged from 1.0~ns to 14~ns, corresponding to 0.25~m to 5.0~m 
and the signals were split and combined into beams
using 8-way Mini-Circuits ZCSC-8-13-S+ power splitters.
The system architecture allowed simultaneous digitization 
of both the antenna channels and the beams. 

  \begin{figure*}[!htbp]
      \begin{center}
        \includegraphics[width=10cm]{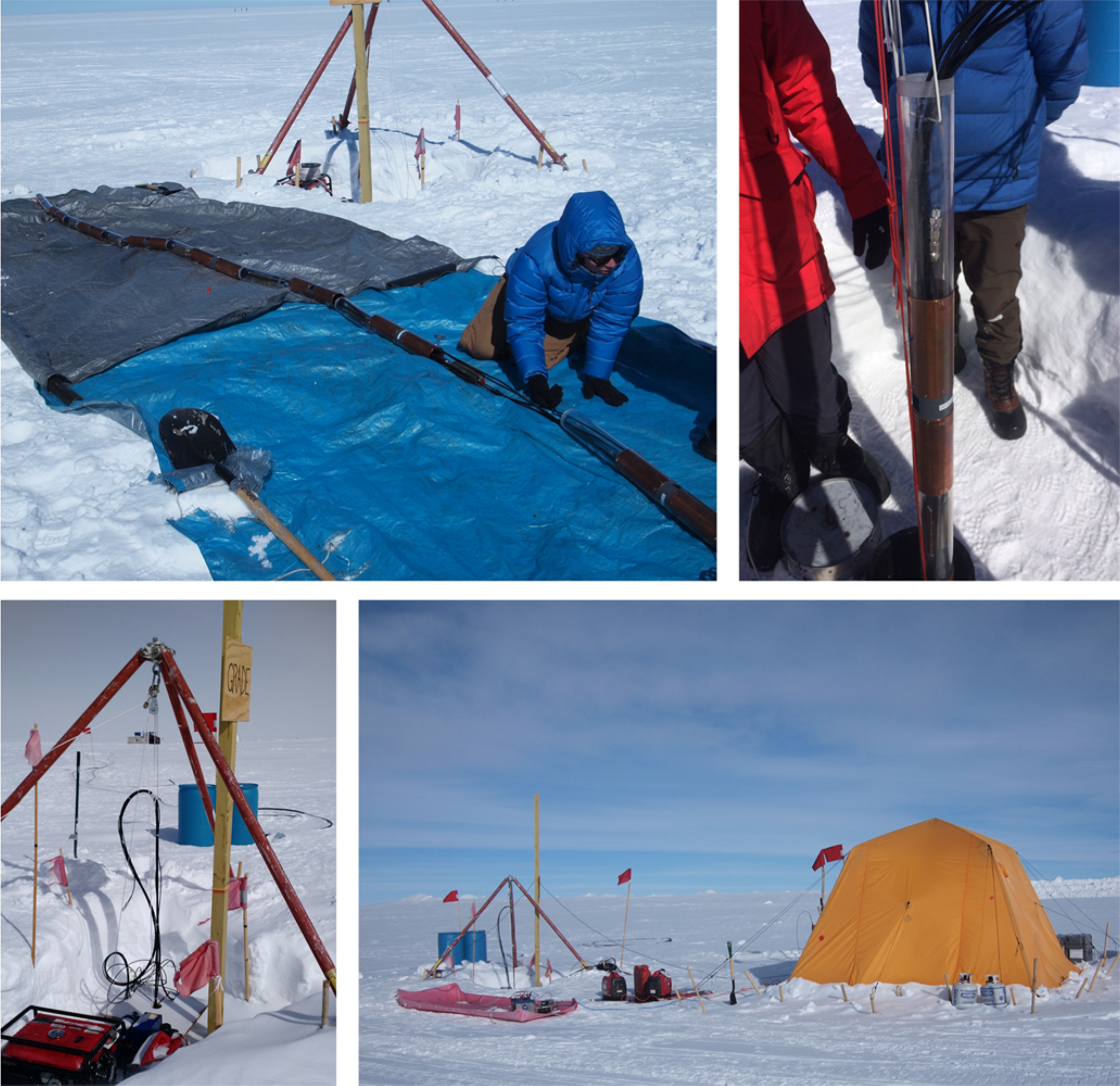}
    \end{center}
    \caption{Pictures of the system deployment in Greenland.  Upper left: laying out the antennas on the surface, 
    before lowering the front-end system down the borehole.  Upper right: lowering the antennas, amplifiers, and 
    cabling down the borehole. Lower left: the system installed down the borehole. Lower right: the deployment
    camp, including deployment gantry and data-taking tent.} 
    \label{fig:disc}
  \end{figure*}  

The instrument included up to eight broadband dipole antennas, described in
Section~\ref{sec:antennas}. The number of antennas included in the array during the 2015 deployment season was 
tunable. 

Waveforms from both branches of the signal chain (antenna channels and beam channels) 
were sampled at 2~GSa/s with record lengths of 1024 points 
with four DRS4 evaluation boards\footnote{https://www.psi.ch/drs/evaluation-board}. 
A Tektronix MSO5204B oscilloscope generated a global trigger for the digitizers using 
either automatic (unbiased) triggers or a threshold crossing on either an antenna or beam channel.

The eight antennas and front-end amplifiers were lowered down the DISC 
borehole\footnote{http://www.icedrill.org/equipment/disc.shtml} using the gantry and winch 
system shown in Figure~\ref{fig:disc}.  The DISC borehole is located approximately $0.5$~km from the 
main activities of the station.  A fixed feed-to-feed antenna spacing of 76~cm was achieved using 
plastic spacers. Eight 115~m cables were bundled together at the surface of the ice and fed through 
the center of the topmost antennas and spacers such that signals from each individual antenna could be 
recorded. The bottom of the array was lowered to a depth of 96~m.

During the June 2015 deployment, Summit Station was relatively radio quiet, with two notable exceptions. 
A balloon-borne radiosonde launched daily at Summit Station in support of the ICECAPS atmospheric monitoring 
program appears in our instrument at 402.8 MHz. The signal remains above thermal noise for several minutes, 
and the power falls off as the balloon drifts away. 
Intermittent, transient signals at 150~MHz and 433~MHz also appear 
in the downhole antennas, due to station communications at those frequencies.


\subsection{Noise Correlations at Summit Station}

\begin{figure}[h]
      \begin{center}
        \includegraphics[width=9cm]{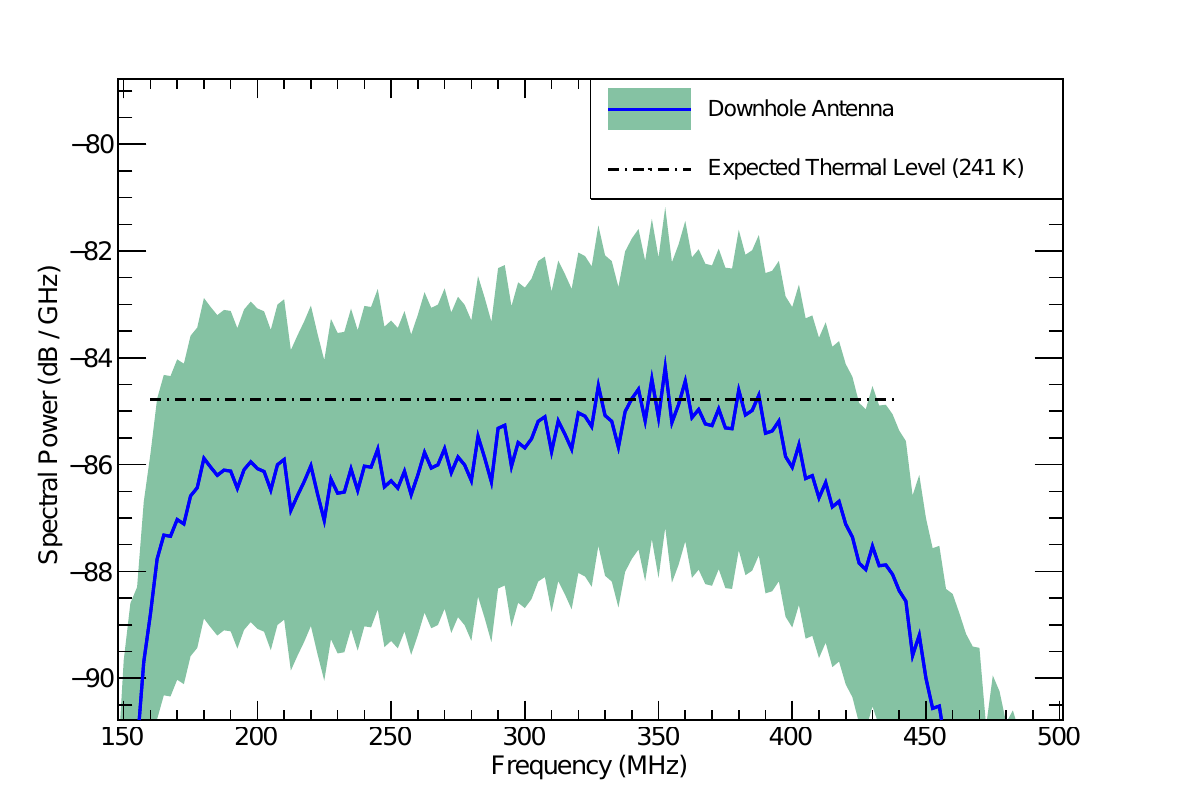}
    \end{center}
    \caption{The average power spectral density for unbiased, thermal-noise 
      triggers of a downhole antenna (blue solid line) compared with 241~K thermal noise (dashed line). A $\pm3$~dB
      error is shown, due to uncertainties in the temperatures of in-ice amplifiers and cables.} 
    \label{fig:noiseSummit}
  \end{figure}

Figure~\ref{fig:noiseSummit} shows the power spectrum measured by a single downhole antenna. The power spectrum 
shown is the mean of 200~individual power spectra of 500-sample waveforms recorded on the oscilloscope
and is corrected for the overall gain in the signal chain.  
Uncertainties in the temperatures of the in-ice amplifiers and cables lead to 
a $\pm3$~dB uncertainty in the spectral power. As shown in the figure, noise observed on the downhole antenna 
is consistent with the expected thermal noise power in the 160-440 MHz band, defined by the 3~dB point
of the filters used in the system, at 241~K, the temperature of the ice at 100~m depth~\cite{grip}.

  \begin{figure*}[!htbp]
      \begin{center}
        \includegraphics[width=14cm]{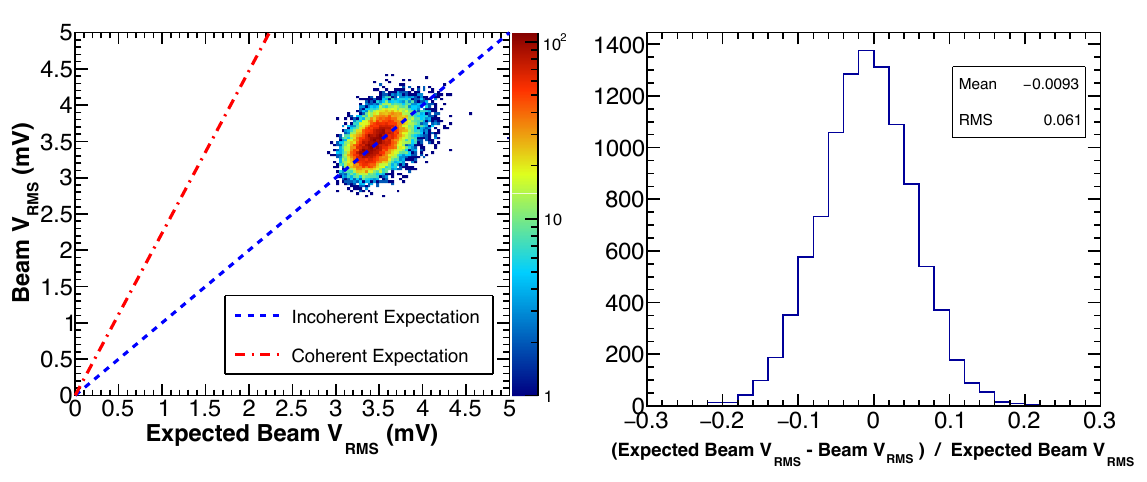}
    \end{center}
    \caption{Expected $V_{\text{RMS}}$ from a beam formed with five antennas compared to the 
      measured $V_{\text{RMS}}$ (left) and the fractional difference between the two (right).  
      The width of the distribution is dominated by the non-Gaussian behavior of the digitizer noise.} 
    \label{fig:noiseRMSSummit}
  \end{figure*}
For uncorrelated noise, the RMS voltage in the beamformed channel, $V_{\text{RMS}, \text{beam}}$, is the quadrature 
sum of the RMS voltages, $V_{\text{RMS},i}$, in the individual antennas after accounting for the relative loss
in the beamformer, $L$, which is -19.9~dB in power 
for the system deployed in Greenland.  The loss in the beamformer 
is dominated by the 8-way splitter and 8-way combiner in each channel.  
Although digitizer noise is small ($<2\%$ for beam channels and $<0.5\%$ for antenna channels)
compared to the system noise (thermal noise plus amplifier noise) level, 
we account for the RMS voltage from digitizer noise in both the
antenna channels, $V_{\mathrm{RMS_D},i},$ and the beam channel, $V_{\mathrm{RMS_D, \mathrm{beam}}}$.
The expected $V_{\text{RMS, beam}}$ is given by

\begin{equation}
 V_{\text{RMS,beam}}^2 = 10^{L/10} \Big( \sum_{\mathrm{N_{antennas}}}\big( V_{\text{RMS},i}^2 - V_{\mathrm{RMS_D},i}^2 \big) \Big) + V_{\mathrm{RMS_D, \mathrm{beam}}}^2.
 \label{eqn:noisecorr}
\end{equation}

Figure~\ref{fig:noiseRMSSummit} compares the expected to the measured  $V_{\text{RMS,beam}}$ for five antennas combined 
using the analog beamformer. This data set consists of 
events recorded at a constant rate (unbiased) over the course of one 
night.  A small fraction ($<0.5\%$) 
of events that have high power in the 160-440~MHz band, indicative of man-made noise, 
are removed from the data set, leaving events 
dominated by thermal noise.  The blue dashed line shows the expectation for an incoherent signal, such as thermal noise, 
while the red dashed line shows the expectation for a coherent signal, such as a true plane wave signal incident on the 
detector.
The $<10\%$ difference between the expected and measured $V_{\text{RMS,beam}}$ shows that 
the noise observed by the downhole antennas adds incoherently when forming a beam, 
indicating that noise from adjacent in-ice antennas is uncorrelated.  This is consistent with our results from
anechoic chamber measurements (see Section~\ref{sec:ncmeas}).
  
\section{Conclusions}
\label{sec:conclusions}
The in-ice phased array for detection of high-energy neutrinos is a promising technique.  
We have performed studies that demonstrate that thermal noise is uncorrelated between tightly-packed
neighboring antennas in each other's nulls both in an anechoic chamber and with an in-ice phased array at 
Summit Station. In the anechoic chamber, we have also validated the beamforming technique through simulations and 
measurements, achieving the expected improvement in SNR when beamforming using an experimental system.
Measurements at Summit Station of the noise environment and ice characteristics
indicate that the site is suitable for an in-ice radio neutrino telescope.

A simulation including realistic parameters for an FPGA-based correlation trigger indicates 
that phasing signals from multiple low-gain antennas yields significant improvement in achieved trigger threshold, 
consistent with idealized expectations and anechoic chamber measurements.  
We are designing and constructing
such a trigger to be implemented first at the South Pole on the ARA experiment. Pending successful demonstration
of the technique, a larger array with hundreds of antennas per station could be proposed either at the South Pole or
in Greenland to achieve a low energy threshold capable of providing significant overlap in energy with IceCube in 
the PeV energy range, and extending the measurement of high energy neutrinos through the higher-energy 
cosmogenic neutrino range.

We would like to thank CH2M Hill and the US National 
Science Foundation (NSF) for the dedicated, knowledgeable, and
extremely helpful logistical support team enabling us to perform our work at Summit Station, 
particularly to J. Jenkins. We are deeply indebted to those who dedicate their
careers to help make our science possible in such remote environments.
We would like to thank the University of Wisconsin-Madison IceCube and 
ARA groups for allowing us to use their anechoic
chamber and the ARA collaboration for lending the ARA bicone and slot antennas for testing.
We also thank D. Arakaki for the use of anechoic chambers at the California Polytechnic State 
University for antenna characterization measurements.
This work was supported by the Kavli Institute for Cosmological Physics at the University of
Chicago, Department of Energy Grant DE-SC0009937, and the Leverhulme Trust. 
Computing resources were provided by the University of Chicago Research Computing Center.

\bibliographystyle{elsarticle-num} 
\bibliography{paper}

\end{document}